\documentclass[runningheads]{llncs}
\usepackage[T1]{fontenc}

\usepackage{subcaption}
\usepackage{graphicx}
\usepackage{algorithm}
\usepackage{algorithmic}
\usepackage{xcolor}
\usepackage{amsmath}
\usepackage{braket}
\usepackage{siunitx}
\usepackage{todonotes}
\usepackage[normalem]{ulem}

\DeclareSIUnit\angstrom{\text {Å}}

\begin{document}

\title{K-ADAPT-VQE: Optimizing Molecular Ground State Searches by Chunking Operators}

\titlerunning{K-ADAPT-VQE}

\author{Tatiana A. Bespalova\inst{1, 2}\orcidID{0000-0003-3300-0843} \and
Oumaya Ladhari\inst{1,2}\orcidID{0009-0002-5318-7561} \and
Guido Masella\inst{1}\orcidID{0000-0001-7108-2996}}

\authorrunning{T.Bespalova et al.}

\institute{QPerfect, 23 Rue du Loess, 67000 Strasbourg, France \and
Universit\'e de Strasbourg and CNRS, ISIS (UMR 7006), 67000 Strasbourg, France}

\maketitle              
\begin{abstract}
Classical simulation of molecular systems is limited by exponential scaling, a hurdle quantum algorithms like Variational Quantum Eigensolvers (VQEs) aim to overcome. Although ADAPT-VQE enhances VQEs by dynamically building ans\"atze, it can remain computationally intensive. This work presents K-ADAPT-VQE, which improves efficiency by adding operators in chunks of K at each iteration. Our results from simulating small molecular systems show that K-ADAPT-VQE substantially reduces the total number of VQE iterations and quantum function calls required to achieve chemical accuracy in molecular ground state calculations.

\end{abstract}

\section{Introduction}

Quantum chemistry is one of the most computationally demanding fields in science due to the exponential growth of the Hilbert space with system size. This exponential scaling renders many accurate quantum chemistry calculations intractable for classical computers, particularly for systems exhibiting strong electron correlations. Since molecular systems are fundamentally governed by quantum mechanics, quantum computing presents a natural and highly promising approach for their study~\cite{Feynman1982}.

Although fault-tolerant quantum computers are not yet available, researchers are actively exploring algorithms that could demonstrate a quantum advantage using today's Noisy Intermediate-Scale Quantum (NISQ) devices~\cite{Preskill_NISQ}. Variational Quantum Algorithms (VQAs), such as the Variational Quantum Eigensolver (VQE)~\cite{Peruzzo_2014}, have become prominent in this context. VQEs employ a hybrid quantum classical approach: a parametrized quantum circuit, known as the ansatz, prepares a trial wavefunction on a quantum processor. The energy of this state is then estimated on a quantum processor, while a classical optimizer updates the circuit parameters to minimize the energy. 

However, conventional VQE methods can suffer from several limitations, including inefficient parameter optimization and the presence of numerous local minima in the energy landscape, especially for larger systems. The Adaptive Derivative-Assembled Pseudo-Trotter ansatz Variational Quantum Eigensolver (ADAPT-VQE) ~\cite{ADAPTVQE} was introduced to mitigate some of these limitations. ADAPT-VQE dynamically builds a compact, problem-specific ansatz by iteratively adding operators from a predefined pool, selecting them based on their contribution to lowering the energy gradient.

In this work, we propose several improvements to the ADAPT-VQE algorithm aimed at enhancing its efficiency without compromising accuracy. Our modifications include an optimized subroutine that reduces the operator pool by pruning unnecessary operators and an improved selection process to further minimize the number of addresses to the quantum function. We call our selection process K-ADAPT-VQE. The key difference from the standard procedure is the simultaneous addition of K operators to the ansatz at each iteration. This ``chunking'' strategy significantly reduces the total number of quantum function evaluations required to achieve chemical accuracy.

Our benchmark focuses on gradient-free optimizers, such as COBYLA and CMA-ES which are particularly interesting for NISQ-era hardware where gradient evaluations are costly. We assess the performance of our enhanced ADAPT-VQE implementation across a diverse set of small molecules at various bond lengths, demonstrating consistent improvements in convergence, accuracy, and overall computational resource requirements.

\section{Method}

This section outlines the methodology employed in our K-ADAPT-VQE approach. We begin by describing the preparation of the molecular Hamiltonian and the construction of the operator pool. We then detail the operator selection mechanism, highlighting the K-ADAPT-VQE criterion, and conclude with the full algorithmic procedure.

\subsection{Hamiltonian preparation}
The starting point of the K-ADAPT-VQE algorithm is the formulation of the chemical problem, given by the electronic structure Hamiltonian for a molecular system, typically expressed in the second quantization formalism. The general form of this Hamiltonian is:
\begin{equation}
\hat{H} = \sum_{pq} h_{pq} \hat{a}_p^\dagger \hat{a}_q + \frac{1}{2} \sum_{pqrs} h_{pqrs} \hat{a}_p^\dagger \hat{a}_q^\dagger \hat{a}_r \hat{a}_s,
\end{equation}
where \( \hat{a}_p^\dagger \) and \( \hat{a}_q \) are fermionic creation and annihilation operators acting on spin-orbitals, and the coefficients \( h_{pq} \) and \( h_{pqrs} \) are the one- and two-electron integrals arising from the molecular orbital basis \cite{McArdle2020}. These interactions account for the kinetic energy of the electrons and the Coulomb interactions between them.

To obtain these integrals, we first use PySCF~\cite{sun2018pyscf}, an open-source quantum chemistry package, to perform initial Hartree-Fock (HF) calculations and define the molecular orbitals. The resulting integrals are then processed using OpenFermion~\cite{McClean_2020}, a library tailored for quantum simulations of chemistry, to construct the second-quantized Hamiltonian.

We then map this fermionic Hamiltonian into a qubit Hamiltonian suitable for quantum computation using the Jordan-Wigner (JW) transformation, a standard method that preserves the antisymmetric nature of fermionic wavefunctions. We choose the JW mapping for its conceptual and implementation simplicity, although we note that for other physically (chemically) inspired fermion-to-qubit mappings~\cite{parella2024reducing} the ADAPT-VQE pipeline would remain unchanged aside from the mapping step. 

For small to moderate bond lengths, in the molecular systems studied in this work (e.g. BeH$_2$, LiH, N$_2$) a good approximation to the ground-state can be obtained from restricted HF methods. Conversely, for systems near-degeneracies or stretched bond lengths unrestricted HF may offer better approximations.

\subsection{Operator pool}

The ADAPT-VQE framework relies on an operator pool from which the VQE ansatz is dynamically constructed. In many classical quantum chemistry approaches, couple-cluster (CC) theory is used to capture electron correlation beyond mean-field theory. In quantum computing, this leads to the Unitary Coupled-Cluster (UCC) ansatz~\cite{romero_ucc}, which is compatible with VQE frameworks. In this work, we follow the UCCSD approach, which includes single and double excitations. This can, in principle, be extended to include triple and higher excitations, although the computational cost increases significantly.

The operator pool for the K-ADAPT-VQE used in this work is constructed from UCCSD excitation operators, including all symmetry-allowed double fermionic excitations, which are then mapped to qubit operators and used iteratively in the adaptive ansatz construction.  

The UCC ansatz introduces a unitary form of the CC operator by symmetrizing it $\ket{\Psi_{\text{UCC}}} = e^{\hat{T} - \hat{T}^\dagger} \ket{\Psi_{\text{HF}}}$,
where the CC operator $\hat{T} = \sum_k \hat{T}_k$ is composed of excitation operators, and $\hat{T} - \hat{T}^\dagger$ is anti-Hermitian, ensuring that the exponential is unitary and preserves normalization. Here, \(\hat{T}\) is truncated to include only single ($\hat{T}_1 = \sum_{i,j} \theta_{ij} \hat{a}_i^\dagger \hat{a}_j$) and double ($\hat{T}_2 = \sum_{i,j,k,l} \theta_{ij}^{kl} \hat{a}_i^\dagger \hat{a}_j^\dagger \hat{a}_k \hat{a}_l$) excitations. The exponential is then expanded using the Lie-Trotter formula into a product of operators.

In the JW transformation, these operators have a simple form. For example, double excitation for spin-orbitals $p$, $q$, $r$, and $s$ take the form:
\begin{equation*}
\begin{split}
    \hat{a}_p^\dagger \hat{a}_q^\dagger \hat{a}_r\hat{a}_s - \hat{a}_r^\dagger \hat{a}_s^\dagger \hat{a}_p\hat{a}_q = 2 i (X_p Y_q Y_r Y_s + Y_p X_q Y_r Y_s - Y_p Y_q X_r Y_s - Y_p Y_q Y_r X_s - \\ - Y_p X_q X_r X_s - X_p Y_q X_r X_s + X_p X_q Y_r X_s + X_p X_q X_r Y_s  ) \bigotimes_{k=q+1}^{p-1} \hat{Z}_k  \bigotimes_{l=s+1}^{r-1} \hat{Z}_l.
\end{split}
\label{eq:double}
\end{equation*}
We note that everything here commutes, so the expansion of $e^{\theta \left(\hat{a}_p^\dagger \hat{a}_q^\dagger \hat{a}_r\hat{a}_s - \hat{a}_r^\dagger \hat{a}_s^\dagger \hat{a}_p\hat{a}_q \right)}$ to $\prod_\alpha e^{\pm 2 i \theta \hat P_\alpha}$ is exact. We also note that the double excitations do not act on any qubits between $q$ and $r$.

To enhance computational efficiency, and focus the ansatz construction, when starting from the HF state we exclude from the operator pool: all single excitations (as HF is already optimized for those), all operators that do not conserve the total $z$-projection of spin (as the state we are searching for here is in the symmetry sector of zero spin projection), and we only allow hopping from two occupied in HF orbitals to two unoccupied as the influence of the rest is small on HF. A similar methodology for operator pool construction is used in \cite{ramoa2025reducing}, although in our case we include $\hat{a}_{\uparrow i}^\dagger \hat{a}_{\uparrow j}^\dagger \hat{a}_{\uparrow k}\hat{a}_{\uparrow l}$ and $\hat{a}_{\downarrow i}^\dagger \hat{a}_{\downarrow j}^\dagger \hat{a}_{\downarrow k}\hat{a}_{\downarrow l}$ operators in the pool, but exclude the operators representing electron jumps into already occupied spin-orbitals or from already vacant spin-orbitals.

\subsection{Selecting the operators}

A core component of ADAPT-VQE is the iterative selection of operators based on the largest contribution to the energy gradient. The energy gradient resulting from adding an operator $\hat{P}_\text{sum}$ to the ansatz ($\ket{\psi}$) has a convenient form:
$\frac{\partial E}{\partial \theta} \left. \right\rvert_{\theta=0} = \frac{\partial}{\partial \theta} \bra{\Psi} e^{-i \theta \hat{P}_\text{sum}} H e^{i \theta \hat{P}_\text{sum}}\ket{\Psi} \left. \right\rvert_{\theta=0}=\\ -i\bra{\Psi} e^{ -i \theta \hat{P}_\text{sum}} [\hat{P}_\text{sum}, H] e^{i \theta \hat{P}_\text{sum}}\ket{\Psi}\left. \right\rvert_{\theta=0} = -i \bra{\Psi} [\hat{P}_\text{sum}, H] \ket{\Psi}  $.

To optimize computational cost we precompute the analytical form of the commutators and, at each step of the adaptive procedure, the expectation values of all operator pool commutators. We note that the expectation values for all of the needed Pauli strings could also be precomputed at each step and then combined into the commutator expectations. 

The main addition of the K-ADAPT-VQE to the ADAPT-VQE procedure is in the way operators from the pool are chosen and added to the ansatz. Instead of adding only the operator with the largest energy gradient, we choose the $K$ operators with the highest absolute values of the commutator expectation and add them to the ansatz from the highest gradient to the lowest with the initial value of the parameters equal to zeros. In the following, we use $K=5$, but it is worth noting that it is possible to adjust the $K$ number of operators to the system under study.

\subsection{K-ADAPT-VQE Algorithm}

\begin{itemize}
    \item[\textbf{1.}]\textbf{Prepare the system:}
        First, we compute the molecular integrals using a classical electronic structure package (e.g. PySCF), then we construct the second-quantized fermionic Hamiltonian which is mapped to qubit operators using one of the fermion-to-qubit transformations (e.g. JW)
    
    \item[\textbf{2.}]\textbf{Initialize the ansatz:}
        The first step is to set the initial quantum state \( |\Psi_0 \rangle \) to the HF state. We then start with an empty ansatz.
   
    \item[\textbf{3.}]\textbf{Construct the operator pool:}
        The operator pool includes all symmetry-allowed spin-conserving double excitation operators that correspond to electron hopping from two HF-occupied orbitals to two virtual orbitals. We proceed by mapping all allowed fermionic operators to qubit operators.

    \item[\textbf{4.}]\textbf{Repeat until convergence:}
    \begin{itemize}
        \item[\textbf{a.}] Compute the energy gradient for each operator \( \hat{P}_i\) in the pool.
        \item[\textbf{b.}] Select the top K operators with the largest absolute gradient magnitudes.
        \item[\textbf{c.}] Append these K operators to the ansatz (in order of decreasing gradient).
        \item[\textbf{d.}] Optimize the full ansatz parameters including previously added ones using a classical optimizer (e.g. COBYLA), allowing a maximum number of VQE iterations per ADAPT step (e.g. 200)
        \item[\textbf{e.}] Check convergence: if all gradients are below a threshold \( \epsilon \), stop. Otherwise, return to step 4a.
    \end{itemize}    
    \item[\textbf{5.}]\textbf{Output:}
        We get the final optimized ansatz, parameter set, as well as the ground state energy estimate.
\end{itemize}

\section{Results}

We implemented the K-ADAPT algorithm using the MIMIQ simulator \cite{leonteva2025comparative} to compute the ground state energies for several molecules, with systems up to 20 qubits.
This section presents a comparative analysis of our K-operator chunking strategy (specifically with $K = 5$, termed 5-ADAPT) against  the standard one-operator-at-a-time ADAPT-VQE (termed 1-ADAPT).
All calculations presented utilize the STO-3G basis set and the COBYLA optimizer with a tolerance of $10^{-3}$ Hartree for the VQE subroutine.
The HF and Full Configuration Interaction (FCI) energies, obtained from PySCF, serve as benchmarks.

\begin{figure}[ht!]
    \centering
    \begin{subfigure}[b]{0.517\textwidth}
        \centering
        \includegraphics[width=\linewidth]{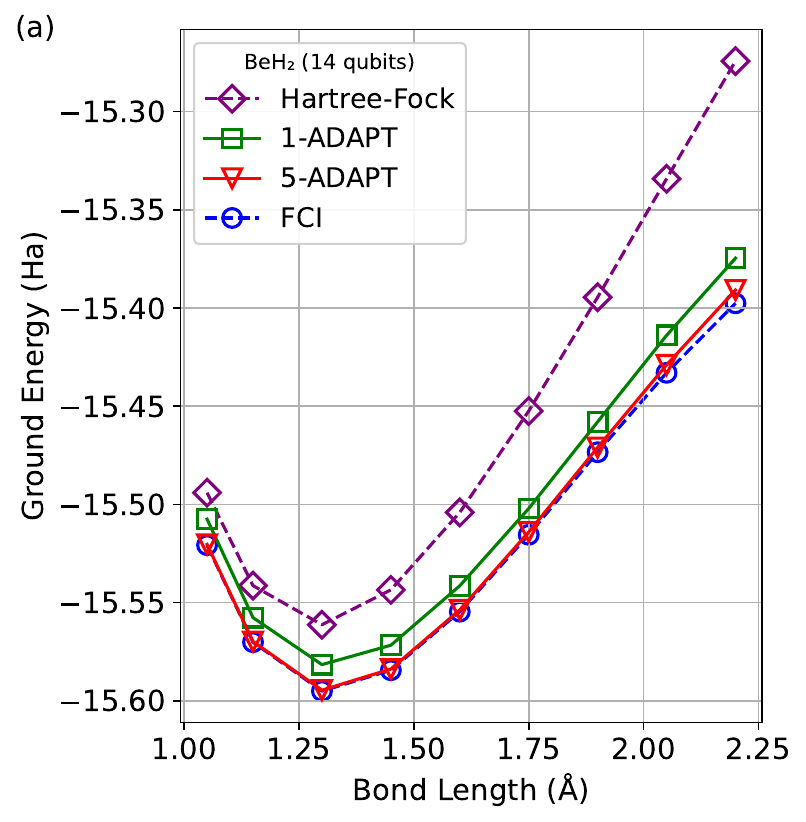}
        \label{fig1}
    \end{subfigure}
    \begin{subfigure}[b]{0.474\textwidth}
        \centering
        \includegraphics[width=\linewidth]{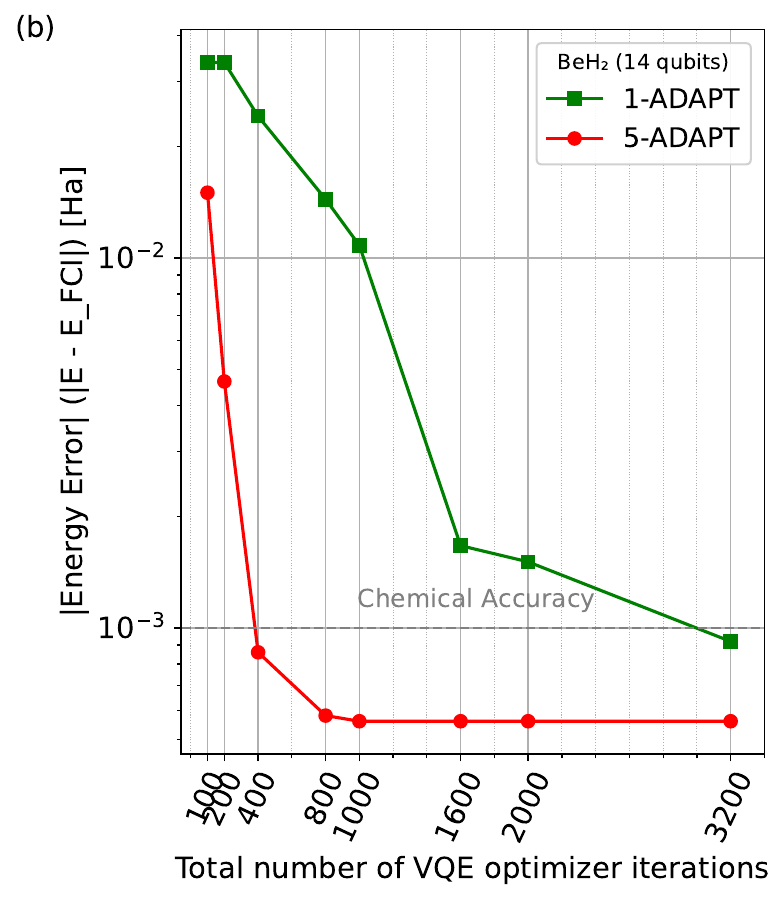}
        \label{fig2}
    \end{subfigure} \\ \vspace{-0.5cm}
    \begin{subfigure}[b]{0.495\textwidth}
        \centering
        \includegraphics[width=\linewidth]{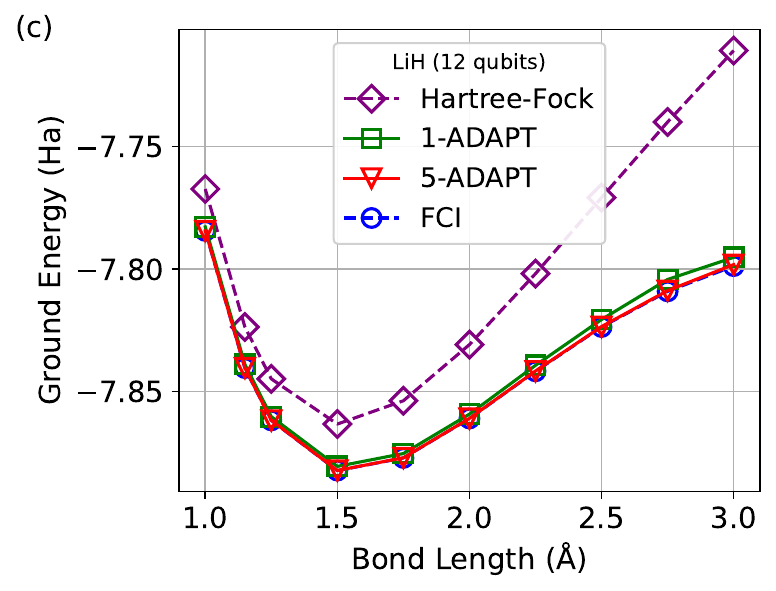}
        \label{fig3}
    \end{subfigure}
    \begin{subfigure}[b]{0.495\textwidth}
        \centering
        \includegraphics[width=\linewidth]{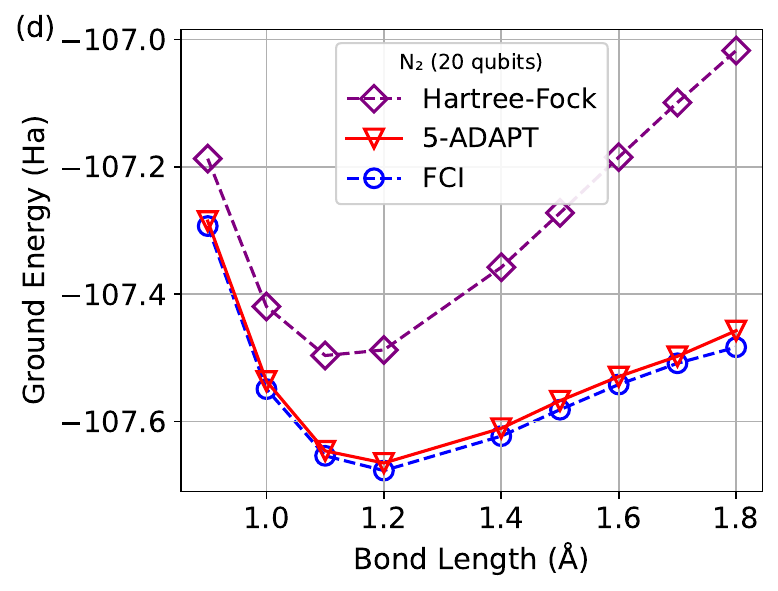}
        \label{fig4}
    \end{subfigure}
    \vspace{-1cm}
    \caption{Performance of 5-ADAPT and 1-ADAPT. All calculations employ the STO-3G basis, the HF and FCI energies serve as benchmarks. In figures a, c and d, the ground state energy (in Hartree) was computed as a function of the bond length for 3 different molecules using ADAPT-VQE with COBYLA optimizer (tolerance 1e-3) and a maximum of 200 VQE iterations per 5-ADAPT step, 40 VQE iterations per 1-ADAPT step. (a) BeH$_2$ molecule (14 qubits): with 25 operators added. (b) Convergence behavior at a fixed bond length for BeH$_2$ of 1.3 $\text{\AA}$, showing the energy error (difference from FCI) versus the number of VQE iterations. (c) LiH molecule (12 qubits): with 25 operators added. (b) N$_2$ molecule (20 qubits): with 50 operators added.}
    \label{fig:combined}
\end{figure}

Figure~\ref{fig:combined} showcases the performance benefits of the K-ADAPT strategy across various molecules and bond lengths.

For the linear BeH$_2$ molecule Fig.~\ref{fig:combined}(a) shows the computed ground state energies as a function of the Be-H bond length.
For both the 5- and 1-ADAPT procedures, the total number of VQE iterations is $1000$ (200 iterations for 5-ADAPT chunks, 40 iterations for each 1-ADAPT operator), and the resulting ansatz consists of 25 operators.
The 5-ADAPT approach consistently yields energies closer to FCI, achieving chemical accuracy (error $< 10^{-3}\,\mathrm{Ha}$) for most of the presented bond lengths. In contrast, the 1-ADAPT procedure, generally results in energies with errors approximately an order of magnitude larger.
The improved convergence of the K-ADAPT method is further detailed in Fig.~\ref{fig:combined}(b), which shows the energy error relative to FCI as a function of the total number of VQE iterations for $\mathrm{Be}\mathrm{H}_2$ at the fixed bond length of $\SI{1.3}{\angstrom}$.

To reach chemical accuracy, the 5-ADAPT strategy required significantly fewer calls to the quantum function. Assuming three evaluations per COBYLA iteration, the 5-ADAPT approach (5 chunks of 5 operators, 200 VQE iterations per chunk for the first 800 iterations, then adapting) reached chemical accuracy with approximately 3300 total quantum function evaluations ($3 \times 800 + 180 \times \frac{25}{5}$). The 1-ADAPT strategy (25 operators added one-by-one, 40 VQE iterations per operator) required roughly 14100 evaluations ($3 \times 3200 + 180 \times 25$). This signifies a computation cost reduction of $\sim 4.3$ times.

We observe that, in the ADAPT-VQE procedure, the parameters associated with the most recently added operators undergo the most significant changes during each VQE optimization. However, allowing previously added parameters to be fine-tuned in subsequent steps is crucial in finding the global energy minima and contributes to the overall efficiency of the optimization.

Figure~\ref{fig:combined}(c, d) further demonstrates the advantages of the 5-ADAPT approach for other molecules. Fig~\ref{fig:combined}(c) shows results for the $\mathrm{Li}\mathrm{H}$ molecule (12 qubits), where 25 operators were added in total. Fig~\ref{fig:combined}(c) shows results for N$_2$ (20 qubits), with 50 operators added. In both cases, the 5-ADAPT method demonstrates robust performance across different bond lengths. 

\begin{figure}[htb]
    \includegraphics[width=\textwidth]{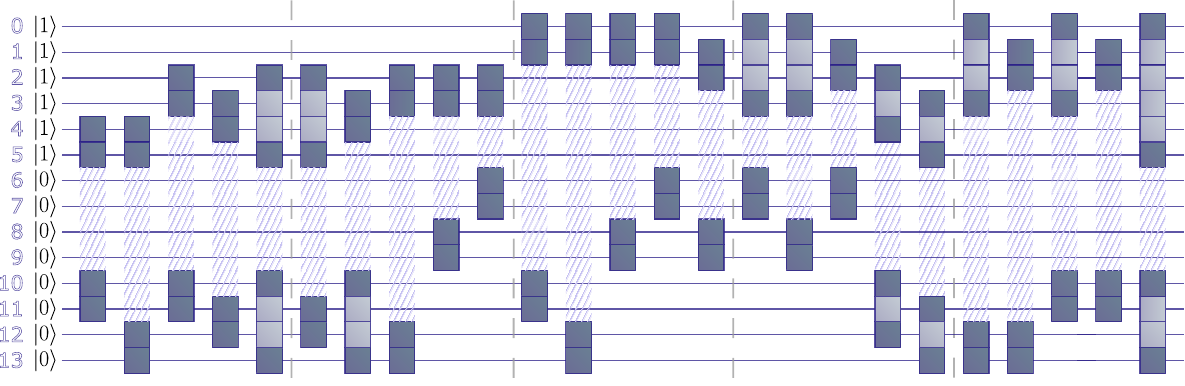}
    \caption{The resulting structure of the ansatz for BeH$_2$ molecule with $1.3 \text{\AA}$ bond length in STO-3G basis. Here each gate denotes $e^{\theta \left(\hat{a}_p^\dagger \hat{a}_q^\dagger \hat{a}_r\hat{a}_s - \hat{a}_r^\dagger \hat{a}_s^\dagger \hat{a}_p\hat{a}_q \right)}$ operator with the variable coefficient $\theta$, each operator is expanded into eight sequential Pauli rotations $e^{\pm 2 i \theta \hat P}$. $p,q,r,s$ correspond to the positions of dark-gray squares, in the Pauli strings $\hat P$ there are $X$ and on these positions, on the positions corresponding to light-gray squares there are $Z$, the rest are $I$.} \label{fig:ansatz}
\end{figure}

In Fig.~\ref{fig:ansatz} we show the resulting ansatz structure for the $\mathrm{Be}\mathrm{H}_2$ molecule at a bond length of $\SI{1.3}{\angstrom}$, generated by the 5-ADAPT algorithm. We observe that operators within the same chunk frequently act on intersecting sets of qubits. This is a consequence of K-ADAPT which selects the top K operators based purely on gradient magnitude, without imposing a disjoint support criterion, distinguishing it from methods like TETRIS-ADAPT-VQE~\cite{anastasiou2024tetris}. We hypothesize that this way of focusing only on the physically relevant operators, even if overlapping, contributes to a more efficient search of the optimal parameters, without increasing the ansatz complexity with less impactful, yet parallelizable operators. 

Furthermore, we note that the presence of spin-conserving excitations within the same spin type, in the fourth chunk of the ansatz for example (Fig.~\ref{fig:ansatz}), indicates their meaningful contribution to the ansatz, and hence supports our choice of including them in the operator pool.

\section{Conclusion}

In this work, we introduced K-ADAPT-VQE, a strategy that enhances the efficiency of the ADAPT-VQE algorithm by adding operators in chunks of K (empirically set to $K=5$ in our demonstrations). This approach significantly reduces overall computational costs, most notably by decreasing the total number of VQE optimization iterations and, consequently, the calls to the quantum function required to reach chemical accuracy. For instance, our 5-ADAPT variant achieved a roughly 4.3-fold reduction in quantum function calls for the $\mathrm{Be}\mathrm{H}_2$ molecule compared to the standard one-operator-at-a-time ADAPT-VQE while attaining comparable accuracy.

The K-ADAPT-VQE method is effective even when operators within a single chunk act on overlapping sets of qubits, a practical scenario that distinguishes it from approaches mandating disjoint supports. This focus on adding the K most gradient-impactful operators, irrespective of their qubit intersection, prioritizes the inclusion of physically relevant correlations, thereby streamlining the ground-state search. These findings suggest that chunking operators is a beneficial strategy for molecular simulations, especially in resource-limited computational environments.

\subsubsection{Acknowledgments.} We thank Rosario Roberto Riso for insightful discussions on quantum chemistry.
This research has received funding from the European
Union’s Horizon 2020 research and innovation programme
under the Marie Skłodowska-Curie Grant Agreement No.
955479 (MOQS - Molecular Quantum Simulations).

\subsubsection{Disclosure of Interests.} Guido Masella is a shareholder of QPerfect.

\bibliography{literature.bib}

\end{document}